\documentstyle[11pt,epsf]{article}
\begin{document}
\catcode`@=11
\long\def\@caption#1[#2]#3{\par\addcontentsline{\csname
  ext@#1\endcsname}{#1}{\protect\numberline{\csname
  the#1\endcsname}{\ignorespaces #2}}\begingroup
    \small
    \@parboxrestore
    \@makecaption{\csname fnum@#1\endcsname}{\ignorespaces #3}\par
  \endgroup}
\catcode`@=12
\def\marginnote#1{}
\newcommand{\newc}{\newcommand}
\newc{\gsim}{\lower.7ex\hbox{$\;\stackrel{\textstyle>}{\sim}\;$}}
\newc{\lsim}{\lower.7ex\hbox{$\;\stackrel{\textstyle<}{\sim}\;$}}
\newc{\mtpole}{M_t}
\newc{\mbpole}{M_b}
\newc{\mqpole}{m_q^{\rm pole}}
\newc{\mlpole}{m_l^{\rm pole}}
\newc{\MS}{{\rm\overline{MS}}}  \newc{\msbar}{\MS}
\newc{\DR}{{\rm\overline{DR}}}
\newc{\tanb}{\tan\beta}
\newc{\stopq}{{\widetilde t}}
\newc{\stopone}{\widetilde t_1}
\newc{\stoptwo}{\widetilde t_2}

\newc{\mstopq}{m_{\tilde t}}
\newc{\mstopl}{m_{\tilde t_L}}
\newc{\mstopr}{m_{\tilde t_R}}
\newc{\mstopone}{m_{\tilde t_1}}
\newc{\mstoptwo}{m_{\tilde t_2}}
\newc{\zbb}{Z\to b\bar}
\newc{\Gb}{\Gamma (Z\to b\bar b)}
\newc{\Gh}{\Gamma (Z\to {\rm hadrons})}
\newc{\xci}{m_{\chi^{\pm}_i}}
\newc{\xcj}{m_{\chi^{\pm}_j}}
\newc{\xck}{m_{\chi^{\pm}_k}}
\newc{\xni}{m_{\chi^{0}_i}}
\newc{\xnj}{m_{\chi^{0}_j}}
\newc{\xnk}{m_{\chi^{0}_k}}

\newc{\sti}{m_{\tilde {t}_i}}
\newc{\stj}{m_{\tilde {t}_j}}
\newc{\stk}{m_{\tilde {t}_k}}
\newc{\sbi}{m_{\tilde {b}_i}}
\newc{\sbj}{m_{\tilde {b}_j}}
\newc{\sbk}{m_{\tilde {b}_k}}

\newc{\cw}{\cos\theta_W}
\newc{\sw}{\sin\theta_W}
\newc{\mhp}{m_{H^\pm}}
\newc{\mA}{m_{A^0}}

\def\NPB#1#2#3{Nucl. Phys. {\bf B#1} (19#2) #3}
\def\PLB#1#2#3{Phys. Lett. {\bf B#1} (19#2) #3}
\def\PLBold#1#2#3{Phys. Lett. {\bf#1B} (19#2) #3}
\def\PRD#1#2#3{Phys. Rev. {\bf D#1} (19#2) #3}
\def\PRL#1#2#3{Phys. Rev. Lett. {\bf#1} (19#2) #3}
\def\PRT#1#2#3{Phys. Rep. {\bf#1} (19#2) #3}
\def\ARAA#1#2#3{Ann. Rev. Astron. Astrophys. {\bf#1} (19#2) #3}
\def\ARNP#1#2#3{Ann. Rev. Nucl. Part. Sci. {\bf#1} (19#2) #3}
\def\MPLA#1#2#3{Mod. Phys. Lett. {\bf A#1} (19#2) #3}
\def\ZPC#1#2#3{Zeit. f\"ur Physik {\bf C#1} (19#2) #3}
\def\APJ#1#2#3{Ap. J. {\bf#1} (19#2) #3}
\def\beq{\begin{equation}}
\def\eeq{\end{equation}}
\def\bea{\begin{eqnarray*}}
\def\eea{\end{eqnarray*}}
\newc{\rb}{R_b}			\newc{\rbmax}{\rb^{\rm max}}
\newc{\mtop}{m_t}		\newc{\mtopmax}{{\mtop}_0}
\newc{\mbot}{m_b}
\newc{\vtb}{\widetilde V_{tb}}
\newc{\ths}{\theta_{\widetilde t}}
\newc{\mz}{m_Z}			\newc{\mw}{m_W}
\newc{\mhalf}{m_{1/2}}
\newc{\mgut}{M_X}
\newc{\ie}{{\it i.e.}}		\newc{\etal}{{\it et al.}}
\newc{\eg}{{\it e.g.}}		\newc{\etc}{{\it etc.}}
\newc{\hpm}{{H^\pm}}		\newc{\mhpm}{m_\hpm}
\newc{\stp}{{\widetilde t}}
\newc{\stl}{{\stp_L}}		\newc{\str}{{\stp_R}}
\newc{\stopl}{\stl}		\newc{\stopr}{\str}
\newc{\mstone}{m_\stone}	\newc{\stone}{{\widetilde t_1}}
\newc{\msttwo}{m_\sttwo} 	\newc{\sttwo}{{\widetilde t_2}}
\newc{\msbotl}{m_{\widetilde b_L}}
\newc{\msbotr}{m_{\widetilde b_R}}
\newc{\msbotone}{m_{\widetilde b_1}}
\newc{\msbottwo}{m_{\widetilde b_2}}
\newc{\mHonesq}{m^2_{H_1}}	\newc{\mHtwosq}{m^2_{H_2}}
\newc{\stau}{{\widetilde\tau}}
\newc{\gev}{{\rm\,GeV}}		\newc{\tev}{{\rm\,TeV}}
\newc{\mev}{{\rm\,MeV}}		\newc{\kev}{{\rm\,keV}}
\newc{\mchone}{m_{\chone}}	\newc{\chone}{\chi_1^\pm}
\newc{\mchtwo}{m_{\chtwo}}	\newc{\chtwo}{\chi_1^\pm}
\newc{\mneone}{m_{\neone}}      \newc{\neone}{\chi_1^0}
\begin{titlepage}
\begin{flushright}
{\large
UM-TH-94-16 \\
hep-ph/9505207\\
April 1995 \\
}
\end{flushright}
\vskip 2cm
\begin{center}
{\large\bf  A Global Fit of LEP/SLC Data with Light Superpartners }
\vskip 1cm
{\Large
G.L. Kane\footnote{{\tt gkane@umich.edu}},
Robin G. Stuart\footnote{{\tt stuart@thoth.physics.lsa.umich.edu}},
James D. Wells\footnote{{\tt jwells@walden.physics.lsa.umich.edu}}
\\}
\vskip 2pt
{\large\it Randall Physics Laboratory, University of Michigan,\\ Ann Arbor,
MI 48109--1120, USA}\\
\end{center}
\vskip 2.0cm
\begin{abstract}
We find that re-analyzing the LEP/SLC data with
light superpartners and low
$\alpha_s(\mz^2)\simeq 0.112$ yields a better fit to the data than the
Standard Model, gives a satisfactory description
of the $R_b$ measurement, and gives a better fit to
$A_{LR}$.
A large body of low energy ($q^2 \ll \mz^2$) data and analyses
provide compelling evidence
for $\alpha_s(\mz^2)\simeq 0.112$.
Global fits to LEP/SLC data in the Standard Model,
however, converge on a value of $\alpha_s(\mz^2)\simeq 0.126$.
Recently it has become increasingly clear that these should be viewed as
incompatible rather than values that can be averaged.
We investigate the possibility that new physics is
causing the LEP high value.  To this end we have conducted a global
analysis of LEP/SLC data in the Standard Model and also in
the Minimal Supersymmetric
Standard Model.
Several predictions could confirm (or rule out)
the results of this paper:  light chargino and stop, top decays into
stop and neutralino, large $R_b$, large $A_{LR}$, and a higher $M_W$.
We briefly discuss the implications of low $\alpha_s$
for more fundamental high-scale supersymmetric theories.

\end{abstract}
\end{titlepage}
\setcounter{footnote}{0}
\setcounter{page}{2}
\setcounter{section}{0}
\setcounter{subsection}{0}
\setcounter{subsubsection}{0}

\baselineskip=0.25in

\section*{Introduction}

Recently it has become increasingly likely that there exists a
genuine and tantalizing
discrepancy between low energy ($q^2 \ll M_Z^2$) determinations
of $\alpha_s$ and the value of $\alpha_s$ extracted from
LEP/SLC data at the $Z$-peak.  Shifman~\cite{shifman95:605}
has argued persuasively that the high value of $\alpha_s(\mz^2)\simeq 0.126$
obtained by fits to $q^2=\mz^2$ data is incompatible with
the values of $\alpha_s(\mz^2)\simeq 0.112$ extracted from low
energy observables
and run up to the $Z$ scale.  Indeed, graphical
demonstrations~\cite{webber94:p15}
of all the various determinations of $\alpha_s$ clearly show
an apparent systematic separation of $\alpha_s(\mz^2)$ between the
low energy data and the $Z$-peak data.

In this letter we will assume as correct the
plethora of extremely precise~\cite{precise} low energy determinations of
$\alpha_s(\mz^2)\simeq 0.112$.
Then the extracted $\alpha_s(\mz^2)$ from LEP/SLC
must either settle to a lower central value with more statistics,
or there is a systematic effect which causes LEP/SLC to fit to
an inaccurately high value of $\alpha_s(\mz^2)$.  Our primary goal
in this letter is to investigate whether $\alpha_s(\mz^2)$
extraction in a supersymmetric model can be substantially lower
than the value of $\alpha_s(\mz^2)$ determined from Standard
Model fitting procedures, thus reconciling low energy and
$Z$-peak determinations of $\alpha_s(\mz^2)$.

One way to think of this is as follows.  The LEP/SLC data has been analyzed
assuming the Standard Model is correct.  If instead light superpartners
exist, then a new analysis of the data is required. All output quantities
will change.  In particular, we find that $\alpha_s(\mz^2)$ is allowed
to decrease by about 0.01; $R_b$ is now more consistent with the
experimental data; agreement with $A_{LR}$ is better;
and in general the global fit
to the data is good.  A number of other authors have also noted that if $R_b$
is explained by new physics, then $\alpha_s$ will
decrease (See for example
Refs.~\cite{shifman95:605,clavelli94:343,erler94:203}).
Before such an argument
can be taken seriously, it is necessary to show that it is quantitatively
large enough and also that it does not contradict other observables such
as left-right asymmetries, forward-backward asymmetries and $M_W$.
We have explicitly demonstrated these features.

\section*{Gauge coupling unification and low $\alpha_s$}

Before continuing further, we should digress on a related question:
Is $\alpha_s(\mz^2)\sim 0.112$ compatible
with simple grand unified theories?  One of the early successes
of supersymmetric grand unified theories was their ability to
unify the gauge couplings (e.g., in $SU(5)$) and {\it predict}
values of $\sin^2\theta_W$ and $\alpha_s(\mz^2)$ which were
in accord with experiment.  As the data and analyses
got better, and the errors several times smaller, most
upper limits on measured $\alpha_s(\mz^2)$ started to drop. Simultaneously,
supersymmetry model builders refined their calculations
and the theoretical lower limits on the predicted $\alpha_s(\mz^2)$ rose.
As it stands today, the lower limit on $\alpha_s(\mz^2)$ is $0.126$
in a simple SUSY GUT theory~\cite{bagger95:443}
(no GUT scale threshold effects, intermediate scales, or
non-renormalizable operator effects) with common scalar and gaugino masses,
and squarks bounded below $1\tev$.
While this lower limit is compatible with the quoted~\cite{lep94:187}
$\alpha_s(\mz^2)$
from LEP/SLC data, it is not compatible with
$\alpha_s(\mz^2)\simeq 0.112$.

An $\alpha_s(\mz^2)$ crisis is actually welcome because it demonstrates
that we can learn about high--scale physics from weak--scale data.
It leads us away from minimal models such as the CMSSM~\cite{kane94:6173}
which assume common scalar masses, common gaugino masses, and precise
gauge coupling unification with a desert between the weak scale
and the unification scale. This minimal constrained supersymmetric
model cannot produce $\alpha_s(\mz^2)$ below 0.126 or $R_b$ above
about $0.2168$; it is a very predictive model.
GUT scale threshold effects and non-renormalizable
operators both modify~\cite{langacker93:4028,langacker95:214}
simple notions of gauge coupling
unification based on a continuous running of beta-functions from the
low scale to the high scale, as do effects at intermediate scales
that do not affect the perturbative unification~\cite{martin95:244}.
As low energy data gets better it
starts to resolve gauge coupling palpitations near the unification
scale.  Several authors~\cite{bagger95:443,chankowski95:23} have used
the lower $\alpha_s(\mz^2)$ values to get insight into the form
of possible supersymmetric GUT theories.  This is in stark contrast to
non-supersymmetric GUTs which have extreme difficulty rectifying
the very large first-order problems of gauge coupling unification and
proton decay constraints with second-order threshold
corrections~\cite{barbieri92:752}, as well as keeping the weak
scale and unification scale naturally separate.

It has been suggested~\cite{roszkowski95:358}
that if one simply abandons the common
gaugino mass assumption then low values of $\alpha_s(\mz^2)$ can be
obtained.  While we fully agree with Ref.~\cite{roszkowski95:358} on the
importance of resolving this $\alpha_s$ ``crisis'', this is a
dramatic approach, and a testable one.  It is disquieting because
in a simple
GUT theory the gauginos must unify in a single adjoint
representation of the GUT gauge group to preserve the gauge symmetry.
If common gaugino masses are discarded then gauge coupling unification
also seems to be gone.  In string theory, however, it is possible to have
gauge coupling unification without having a grand unified group in
four dimensions~\cite{witten85:75}.  Usually it is assumed that the
gauginos will unify as well but this
is not necessarily required.  What is required is the raising of the
unification scale from the typical scale of $10^{16}\gev$ where
simple SUSY theories want to unify, up to the string scale $\sim 10^{18}\gev$.
This is a non-trivial task~\cite{martin95:244}, requiring the introduction
of additional states which affect the running of the gauge couplings.
For these reasons, results based on simple GUT gauge coupling unification
without gaugino mass unification are difficult to obtain in a theory.

In this letter it is not our purpose to promote any specific notions
of the GUT scale theory, and we do not attempt to provide any additional
insight into how a more fundamental high-scale SUSY theory could
{\it predict} a low $\alpha_s(\mz^2)$.
We shall focus instead on the low energy
data, and demonstrate how fits to LEP/SLC $Z$-peak observables with light
superpartners could give lower $\alpha_s(\mz^2)$ than fits without
superpartners.  We know that by combining intermediate
scales~\cite{martin95:244}, which do not
hurt perturbative unification, with high scale threshold
effects~\cite{bagger95:443,chankowski95:23} we can
construct a theory with the couplings and spectrum that we find
in this work.

\section*{Extracting $\alpha_s$ in the Standard Model}

The values of $\alpha_s(\mz^2)$ at the $Z$-peak are extracted, mainly,
from two classes of observables: $\Gamma_{\rm had}$ and jet event shapes.
The most important observables in the $\Gamma_{\rm had}$ class
are $\Gamma_Z$, $R_{\rm lept}\equiv \Gamma_{\rm had}/\Gamma_{\rm lept}$,
and $\sigma_{\rm had}$.
The fits for $\alpha_s(\mz^2)$ in the two approaches
yield~\cite{lep94:187,catani94:hepph361,clavelli95:1117},
\bea
\alpha_s(\mz^2) & = & 0.126\pm 0.005~~~{\rm from}~
            \Gamma_{\rm had}~{\rm observables},~{\rm and} \\
\alpha_s(\mz^2)&= &0.119\pm 0.006~~~{\rm from~
              jet~event~shapes}.
\eea
The error in the $\alpha_s(\mz^2)$ determination from $\Gamma_{\rm had}$
observables is statistics limited.  The error associated
with the jet event shape measurements is mostly theoretical, since
the non-perturbative effects of hadronization must be folded into
the perturbative parton level jet correlations.  Furthermore, the
perturbative QCD calculations for the event shape
measurements~\cite{brodsky93:6389,ellis95:223}
are not universally agreed upon, which
compounds the uncertainty.
We therefore cautiously ignore the jet event shape determination,
which are in any case only $1\sigma$ from the low
values, and concentrate on the $\Gamma_{\rm had}$ observables.

In an effort to analyze all observables at LEP simultaneously in
the Standard Model and in the minimal supersymmetric model we
have implemented supersymmetric loop corrections in
{\tt Z0POLE}~\cite{kniehl92:175} and interfaced it with the CERN library
minimizer {\tt MINUIT}~\cite{james75:343} for a complete $\chi^2$ fitter.
The observables that we use in our $\chi^2$ fit are ${\cal O}_i=\Gamma_Z$,
$\sigma_{\rm had}$, $R_b$, $R_c$, $A_{LR}$, $A^b_{FB}$,
$A^c_{FB}$, $R_{\rm lept}\equiv \Gamma_{\rm had}/\Gamma_{\rm lept}$, and
$A^{\rm lept}_{FB}$.
Next we fix the Higgs mass to a low value
consistent with supersymmetry ($m_h=100\gev$), and let {\tt MINUIT}
find the minimum $\chi^2$ for $M_t$ and $\alpha_s(\mz^2)$.
The $\chi^2$ is defined as
\bea
\chi^2& = &\sum_i \frac{({\cal O}_i^{\rm theory}-{\cal O}_i^{\rm expt})^2}
{(\Delta {\cal O}_i^{\rm expt})^2}.
\eea
All the values  of ${\cal O}^{\rm theory}_i$ are calculated within a
specific model and the better the match between theory and data
the lower the $\chi^2$.
Using the Standard Model we find
\bea
M_t & = & 167\pm 15\gev \\
\alpha_s(\mz^2) & = & 0.123\pm 0.005
\eea
as the results of our $\chi^2$ fit to the observables.
These results are consistent with the fits obtained by
the LEP Electroweak Working Group~\cite{lep94:187} corrected
for a light Higgs.

\section*{Extracting a lower $\alpha_s$ in supersymmetry}

Now we set $\alpha_s(\mz^2)$ to a smaller value
(we choose $0.112$) consistent with the numerous low energy
observables, and map out the supersymmetric parameter space which
yields a {\em better} $\chi^2$ with superpartners in loops and
$\alpha_s(\mz^2)=0.112$
fixed than does the
Standard Model, whose $\chi^2$ minimum is
at $M_t=167\gev$ and $\alpha_s(\mz^2)=0.123$.

The idea that light superpartners might resolve the $\alpha_s(\mz^2)$
discrepancy between high scale and low scale data is hinted at
by the large measured value of
$R_b\equiv \Gamma (Z\to \bar b b)/\Gamma (Z\to {\rm had})$ which is
approximately $2.3\sigma$ from the Standard Model prediction.
It was found in Ref.~\cite{wells94:219} that if $m_{\tilde t_1}$
and $m_{\chi^+_1}$ were both less than about $110\gev$ then the
discrepancy between theory and data for this one observable could
go away.  Since $R_b$ had the highest ``pull'' on the Standard Model
$\chi^2$ for LEP data, resolving this $2.3\sigma$ deviation could
substantially improve the global fit.

If the theoretical prediction
for $R_b$ is raised by increasing the $\Gamma_{\bar bb}$
partial width, then for a fixed $\alpha_s$
the total hadronic decay width is also increased.
To a good approximation the hadronic width of the $Z$ is separable
into an electroweak piece and a QCD correction:
\bea
\Gamma^{\rm theory}_{\rm had}& =&\Gamma^{\rm theory}_{EW,{\rm had}}
\left( 1+\frac{\alpha_s (\mz^2)}{\pi}+\cdots \right)
\Longleftrightarrow  \Gamma^{\rm expt}_{\rm had}
\eea
Although $R_b$ is rather insensitive to the QCD corrections,
the partial widths $\Gamma_{\bar bb}$ and $\Gamma_{\rm had}$ are quite
sensitive.
It is clear from the above equation that if we obtain a higher
$\Gamma^{\rm theory}_{EW,{\rm had}}$ in supersymmetry than was found
in the Standard Model then the QCD corrections must be smaller
in the supersymmetric theory to match the experimental
determination of $\Gamma^{\rm expt}_{\rm had}$; that is, $\alpha_s (\mz^2)$
must be lowered to best fit the data.
Therefore, it qualitatively
appears that we can simultaneously increase $R_b$ and lower
$\alpha_s$, while at the same time keeping $\Gamma^{\rm theory}_{\rm had}$
fixed.

Our next step then is to hone in
on the region of supersymmetric parameter space which will substantially
increase $R_b$~\cite{wells94:219} and check to see that the
$\chi^2$ fit to LEP/SLC data is consistent with low $\alpha_s(\mz^2)$
and {\it all other observables} such as $A_{FB}$, $\Gamma_Z$,
$R_{\rm lept}$, etc.  With light superpartners having a large
effect on observables such as $R_b$, one would expect {\it a priori}
that these same superpartners will affect other observables
at LEP and potentially could yield a worse $\chi^2$ fit to the
data than the Standard Model.  It is imperative that all
observables be analyzed simultaneously to confidently state
that a lower $\alpha_s$ extraction at LEP is possible in supersymmetry.
To be precise about our procedure, we have fixed $\alpha_s (\mz^2)=0.112$
and searched through the MSSM parameter space for
solutions which yield {\it better} $\chi^2$, at fixed $\alpha_s(\mz^2)$,
than the lowest $\chi^2$
fit in the Standard Model where $\alpha_s(\mz^2)$ was allowed to
vary to its best-fit minimum value of 0.123.

We have fixed
$\alpha_s(\mz^2)=0.112$ for two reasons.  One, we want to see if
$\chi^2_{SUSY}$ at a low value of $\alpha_s(\mz^2)\simeq 0.11$ can give
a better $\chi^2$ than the Standard Model.  And, we have determined that
$\alpha_s(\mz^2)=0.112$ is near the best minimum
$\chi^2_{SUSY}$ in this analysis (with heavy first and second generation
squarks and sleptons).  Due to the extremely complicated minimization
procedure with all the free MSSM parameters we do not yet
claim with certainty that
the global minimum of the $\chi^2_{SUSY}$ fit is at $\alpha_s(\mz^2)=0.112$,
but only that there are
at least local minima with $\alpha_s(\mz^2)=0.112\pm0.004$
and $\chi^2_{SUSY}<\chi^2_{SM}$.
Furthermore, we have fixed $\tan\beta$
at its lowest possible
value, which is determined by the top Yukawa remaining
perturbative below the GUT scale,
since this value gives
the best $\chi^2_{SUSY}$ in the region of $\tan\beta <30$.
For $\tan\beta >30$ the light pseudo-scalar Higgs can become
important and we have not yet incorporated it into {\tt Z0POLE}.

We have included into {\tt Z0POLE} all vector boson self-energy diagrams
and vertex corrections which involve the charginos, neutralinos,
stops and sbottoms.  The only light squark or slepton expected in
the spectrum which will affect our analysis
is the $\tilde t_R$, which becomes light through mixing in
the stop mass matrix.
Since the sbottoms are isospin partners to the stops they
must be explicitly included in the calculation.
We expect and assume that all other sparticles have masses too large
to have a significant
impact on the final answer. Although we work basically in a minimal
supersymmetric theory, our results are largely independent of the gluino
mass, and of first and second family squark masses if they are at all
heavy.  Results do assume $M_1=M_2$ (bino and wino masses) at the
GUT scale.  Other
parameters are varied over allowed values (rather than guessed), to
give the regions in the figures.

Our calculations of the one-loop diagrams
were checked in {\tt Z0POLE} by exact numerical cancellations
of the $\log (\mu^2)$
which accompany all divergences in counter terms
of the on-shell renormalization
scheme. These exact cancellations of the $\log (\mu^2 )$ in
all observables and $\Delta r$ are crucial requirements for a
trustworthy calculation.

Figure~\ref{tot}
\begin{figure}
\centering
\epsfxsize=4in
\hspace*{0in}
\caption{Region of supersymmetric parameter space with a better
$\chi^2$ fit with $\alpha_s(\mz^2)=0.112$ than the best standard
model $\chi^2$ fit which was at $\alpha_s(\mz^2)=0.123$.}
\label{tot}
\end{figure}
is a summary of the main result in this letter.  The enclosed
area in the $m_{\chi^+_1}$--$m_{\tilde t_1}$ plane
is the region of parameter space which yields a better
$\chi^2$ fit to LEP/SLC data using supersymmetry and $\alpha_s(\mz^2)=0.112$
than the absolute lowest $\chi^2$ obtained in the Standard Model
(with $\alpha_s(\mz^2)=0.123$).  The SUSY $\chi^2/{\rm d.o.f}$ are
as much as 1/3 better than the Standard Model best fit, and this minimum
occurs when the chargino is near $80\gev$ and the stop is near $60\gev$.
Interestingly, the lower bound on the lightest chargino is about
$58\gev$ although high $R_b$ values were obtained for $m_{\chi^+_1}<58\gev$.
The reasons for this are clear.  The lightest neutralino in this region
of parameter space is too light, and the $Z$ decay width
becomes too large.  The truncated section in the lower right corner
has a straightforward explanation as well.  Here the stop
is always lighter than the lightest neutralino and therefore
becomes the LSP, which we exclude.

It is very interesting to see the effect of
supersymmetry on other observables.   In Figure~\ref{four}
\begin{figure}
\centering
\epsfxsize=4in
\hspace*{0in}
\caption{Four observables versus the lightest chargino mass.  The dotted
line is the measured central value of the observable, and the dashed lines
are the $1\sigma$ limits.  The solid straight line is the Standard Model
best fit value obtained from {\tt Z0POLE} with $m_h=100\gev$,
and the shaded region
that which yields $\chi_{SUSY}^2<\chi^2_{SM}$ as other parameters are varied.
As expected in a better $\chi^2$ fit,
the $R_b$ and $A_{LR}$ predictions fit the experimental values as
measured by LEP/SLC better than the SM does. Note also
that the $W$ mass prediction in supersymmetry is
higher than the Standard Model prediction.  And, the top is expected
to decay into the lightest stop and light neutralinos with branching
fraction as high as $60\%$.  }
\label{four}
\end{figure}
we plot three observables, $R_b$, $M_W$,
and $A_{LR}$ versus
the lightest chargino mass.  The dotted line in each graph is the
central measured value of each of these observables, and the dashed
lines are the $1\sigma$ errors associated with the measurements.
The measured value for $R_b$ is taken from Ref.~\cite{lep94:187},
$M_W$ from~\cite{cdfmw}, and $A_{LR}$
from~\cite{slacalr}.  The solid straight line is {\tt Z0POLE}'s best fit
Standard Model value with $m_h$ fixed at $100\gev$ (the Standard Model
values would disagree more with experiment if $m_h\gsim 300\gev$).
The shaded region
is the range of values obtained (versus lightest chargino mass), as
other parameters vary, which
yield a better $\chi^2$ with light superpartners and $\alpha_s(\mz^2)=0.112$
than the best $\chi^2$ in the Standard Model.

Several aspects of Figure~\ref{four} are important.  The $R_b$ region
is significantly higher than in the Standard Model.  $M_W$ is also
higher.  It is amusing that earlier values of $M_W$ would have
preferred the Standard Model to supersymmetry, but the new
value~\cite{cdfmw} ($80.33\pm 0.17\gev$) does not.  The SUSY
$A_{LR}$ value is closer to the SLC $A_{LR}$ measurement.  These
results translate to $\sin^2\theta_W=0.2312\pm 0.0004$.
The values of $M_t$ that we found
with $\chi^2_{SUSY}<\chi^2_{SM}$ range between $162\gev$ and
$190\gev$.  The upper limit on $M_t$ comes about mostly from the
inability to get low $\tan\beta$ and high $M_t$ simultaneously, and
still keep the top Yukawa perturbative at the high scale. With very
light charginos we run the risk of having top decays into the lightest
stop and light neutralinos be too numerous to be consistent with top quark
production and decay data at Fermilab~\cite{cdftop}.  Figure~\ref{four}
shows that the branching fraction of these supersymmetric top decays can
be as high as $60\%$, and in general much of the parameter space has
a significant top decay branching fraction into supersymmetric states
which could be detected when many more top events are detected
at a high luminosity collider.

It should be re-emphasized that the most important
phenomenological implication of lowering the
extracted $\alpha_s(\mz^2)$ is light superpartners.  Most of
the allowed parameter space in Figure 1 will be detectable at
LEP~II and an upgraded FNAL collider.  With sufficient luminosity
LEP~II will be able to
detect all charginos and stops with masses to within a few GeV of $\sqrt{s}/2$.
An upgraded Tevatron collider should be able to reach
charginos and stops with considerably higher masses~\cite{mrenna95:14}
than LEP.
However, FNAL, and to a limited extent LEP,
has some difficulty
cleanly detecting
a signal for Higgsino-like charginos.  In the limit of pure
Higgsino the LSP mass gets closer and closer to the lightest
chargino mass.  When the chargino decays into LSP plus leptons,
the leptons may have too little energy to trigger on,
so the signal is reduced.  This region of chargino
parameter space is largely the region we are in.

\section*{Conclusion}

We have demonstrated that the extracted value
of $\alpha_s(\mz^2 )$ from LEP/SLC data can be lowered to agree
with other $\alpha_s(\mz^2)$ determinations when
superpartners are added to the fit.  An essential aspect of
this work is the inclusion of all relevant LEP/SLC data, so that
the results are known to be consistent with all observables.  We have
found that light charginos and stops (with masses below $\sim100\gev$)
are required if the total $\chi^2_{SUSY}$ with $\alpha_s(\mz^2)=0.112$
is better than the $\chi^2_{SM}$ with $\alpha_s(\mz^2)$ at its
Standard Model best-fit value of $0.123$.  Our approach is
largely independent of SUSY assumptions.

The SUSY spectrum and couplings required to obtain our results
cannot be obtained
in a fully minimal supersymmetric model.  They can be obtained by
adding the effects of high scale thresholds, and/or Planck scale
operators, and/or perturbatively valid intermediate scales.  It is
very encouraging that data at the electroweak scale seems to be telling
us about physics near the Planck scale.

The resultant supersymmetry parameter space has several
important phenomenological implications:   The $W$ mass is
higher than the expected Standard Model best fit.
$R_b$ and $A_{LR}$ should also be larger than their Standard Model
values.  Light superpartners below about $100\gev$ must exist.
LEP~II and FNAL will probably find these superpartners if they
are this light; if they don't, very precise determinations of
the $W$ mass, $R_b$, or $A_{LR}$ could rule out or further support
this exciting possibility.

\section*{Acknowledgements}

G.K. would like to thank A. El-Khadra, F. Wilczek, and L. Dixon
for emphasizing the validity of the lower $\alpha_s$ determinations
and forcing him to think about them.  R.S. thanks A. Blondel
for helpful correspondences, and B. Kniehl for providing updated
QCD libraries for {\tt Z0POLE}. And J.W. would like to thank C. Kolda
and S. Martin for useful conversations.  This work was supported in
part by the U.S. Department of Energy.

\vfill\eject

\end{document}